\documentclass[prl,amsmath,amssymb,reprint,superscriptaddress]{revtex4-2}

\usepackage{graphicx}
\usepackage{dcolumn}
\usepackage{bm}
\usepackage[utf8]{inputenc}
\usepackage[T1]{fontenc}
\usepackage{mathptmx}
\usepackage{etoolbox}
\usepackage{xcolor}
\usepackage[english]{babel}
\usepackage[colorinlistoftodos, color=green!40, prependcaption]{todonotes}
\usepackage{amsthm}
\usepackage{mathtools}
\usepackage{physics}
\usepackage{adjustbox}
\usepackage{placeins}
\setcitestyle{square,sort,comma,numbers}
\usepackage{microtype}
\usepackage[colorlinks=true,
            linkcolor=blue,
            urlcolor=blue,
            citecolor=blue]{hyperref}
\usepackage[normalem]{ulem}
\usepackage[capitalize]{cleveref}

\DeclareMathAlphabet{\mathcal}{OMS}{cmsy}{m}{n}
 
\definecolor{Lightgray}{rgb}{0.88, 0.88, 0.88}

\makeatletter
\def\@email#1#2{%
 \endgroup
 \patchcmd{\titleblock@produce}
  {\frontmatter@RRAPformat}
  {\frontmatter@RRAPformat{\produce@RRAP{*#1\href{mailto:#2}{#2}}}
  \frontmatter@RRAPformat}
  {}{}
}%
\makeatother
\begin{document}

\preprint{AIP/123-QED}

\title[Intermittent dynamics and NMR]
{Intermittent molecular motion and first passage statistics for the NMR relaxation of confined water}

\author{Simon Gravelle}
\affiliation{Institute for Computational Physics, 
University of Stuttgart, 70569 Stuttgart, Germany}
\affiliation{Univ. Grenoble Alpes, CNRS, LIPhy, 38000 Grenoble, France}
\author{Benoit Coasne}
\affiliation{Univ. Grenoble Alpes, CNRS, LIPhy, 38000 Grenoble, France}
\affiliation{Institut Laue Langevin, 38000 Grenoble, France}
\author{Christian Holm}
\affiliation{Institute for Computational Physics, 
University of Stuttgart, 70569 Stuttgart, Germany}
\author{Alexander Schlaich}
\affiliation{Institute for Atomistic Modeling of Materials in Aqueous Media,
  Hamburg University of Technology,
  21073 Hamburg, Germany}
\email{simon.gravelle@cnrs.fr, alexander.schlaich@tuhh.de}

\begin{abstract}
\noindent 
The structure and dynamics of fluids confined in nanoporous media differ from those in bulk,
which can be probed using NMR relaxation measurements.
We here show, using atomistic molecular dynamics simulations of water in a slit nanopore,
that the behavior of the NMR relaxation rate, $R_1$, with varying surface interaction and confinement strength can be estimated from the exchange statistics of fluid molecules between the adsorbed surface layer and the bulk region, where molecules undergo intermittent dynamics.
We employ first return passage time calculations to quantify the molecular exchange statistics, thereby linking microscopic parameters of the confined fluid---such as adsorption time, pore size, and diffusion coefficient---to the NMR relaxation rate.
This approach allows to predict and interpret the molecular relaxation of fluids at interfaces using merely concepts of statistical mechanics and can be generalized to closed and open geometries.
\end{abstract}

\maketitle

\begin{figure*}
\centering
\includegraphics[width=0.85\linewidth]{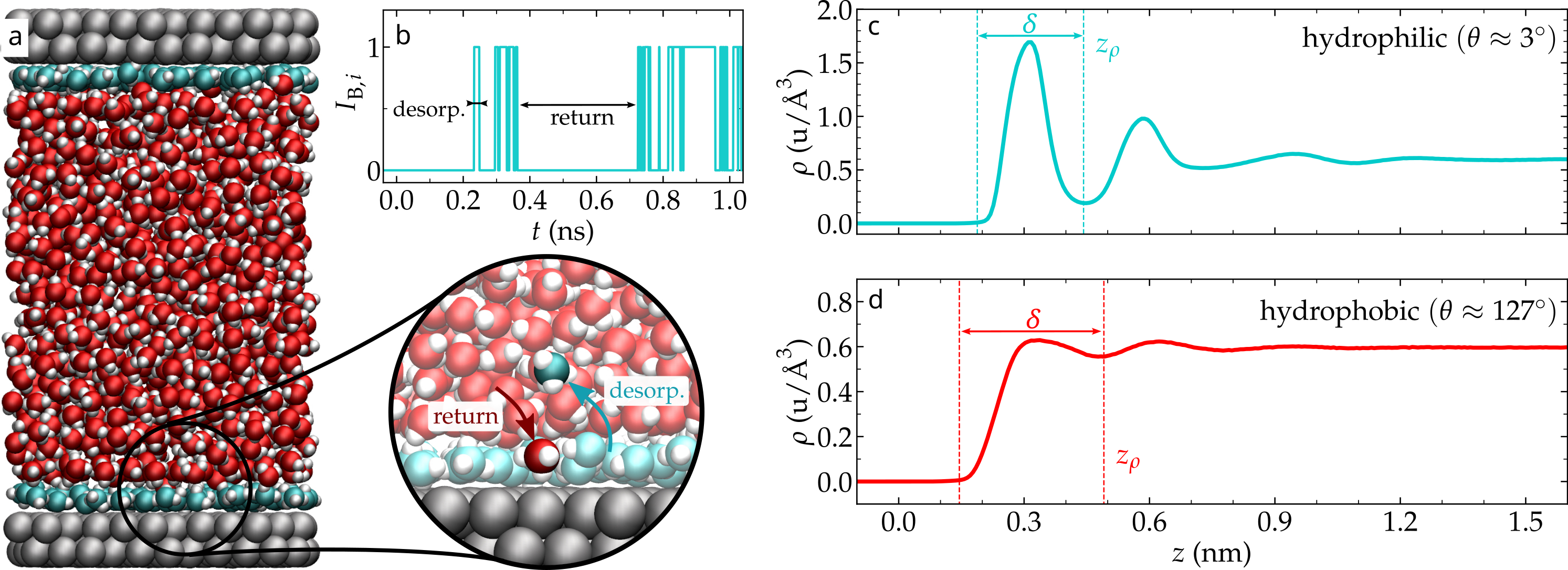}
\caption{
a)~Snapshot of the MD system, where the wall of the nanopore is shown in gray, bulk water molecules in red and white, and surface water molecules in cyan and white.
  The water contact angle is $\theta \approx 127^\circ$~\cite{SI}.
  The inset illustrates the exchange between surface and bulk water, with molecules desorbing from the bottom surface at an average rate $\lambda_\text{desorb}$ and returning to the surface at a rate $\lambda_\text{return}$.
b)~Sample of the indicator function $I_{\text{B}, i}$ as a function of time.
c)~Water density profile for a hydrophilic surface with a contact angle $\theta \approx 3^\circ$. 
   The position $z=0$ corresponds to the position of the first layer of atoms in the wall, $z_\rho$ marks the first minimum in the profile, and $\delta$ represents the width of the surface layer.
d)~Water density profile for a hydrophobic surface with $\theta \approx 127^\circ$.}
\label{fig:system}
\end{figure*}

\noindent Experimental measurements of the nuclear magnetic resonance (NMR) relaxation times are powerful tools for probing the properties of fluids in porous media \cite{ardelean03a}.
For dipolar coupling between two spins, the spin-lattice relaxation time $T_1$
provides insight into the structure and dynamics of confined molecules across a wide frequency range \cite{stapf95a, sattig2014nmr,wardwilliams21a, pinheiro24a}.
NMR relaxometry has been used extensively to characterize liquids in various porous materials \cite{watson97a, kimmich04a, mascotto2017ice}, including silica, zeolites, and calcite \cite{dagostino14a, katsiotis15a, mutisya17a, weigler20182h}.
These relaxometry methods have also been key in elucidating mechanisms of molecular diffusion near solid walls, such as `bulk-mediated surface diffusion', where molecular displacements follow L\'evy-walk statistics \cite{bychuk95a, stapf95a, zavada99a, kimmich02a, levitz05a}.
When a fluid is confined within a porous medium, its molecules can be described as alternating between two states (\cref{fig:system}a): adsorption at the interface and excursions within the bulk of the material \cite{kimmich93a, bychuk95a, levitz19a}.
The average duration that molecules remain adsorbed at the solid surface depends on several factors including the strength of interactions between the solid and liquid phases and the temperature \cite{karger16a, coasne16a, korb18a, bousige21a}.
In contrast, the average time that molecules spend diffusing freely in the bulk---i.e., the interval between desorption from and re-adsorption to the surface---is influenced by parameters such as diffusivity, pore geometry, and the availability of surface adsorption sites. 

These distinct timescales can be related to $T_1$ since the typical exchange rate $W$ between the surface and the bulk is of the order of several GHz, 
which is significantly higher than the $^1$H-NMR relaxation rate ${\cal R}_1$ ($= 1/{\cal T}_1$) for bulk water at ambient conditions (approximately 1\,Hz \cite{hindman73a,paschek24a-pre}). 
In this high-frequency limit $W \gg {\cal R}_1$, the NMR relaxation can be split into a bulk $\left({\cal R}_{1, \text{bulk}}\right)$ and a surface $\left({\cal R}_{1, \text{surf}}\right)$ contribution \cite{jaffel06a, korb11a}.
In this Letter, we employ first-passage time calculations to derive an exact expression for the corresponding surface exchange spectral density $J_\text{surf}$ in a slit pore, incorporating all the microscopic parameters that control the intermittent molecular dynamics, including the pore size $H$ and the desorption rate $\lambda_\text{desorb}$. 
We validate these calculations using molecular dynamics (MD) simulations of liquid water confined within a nano slit pore (Fig.\ref{fig:system}a). 
The derived formalism allows us to predict NMR surface relaxation rates across varying pore hydrophilicity, which we compare to the \emph{exact} $^1$H-NMR relaxation rate obtained from analysis of magnetic dipole-dipole correlations.

When the relaxation process is surface-limited, the observed relaxation rate can be split according to
\begin{equation}
N {\cal R}_1 (\omega) = N_\text{surf} {\cal R}_{1, \text{surf}} (\omega) + N_\text{bulk} {\cal R}_{1, \text{bulk}} (\omega),
\label{eq:Ntot_Nsurf_Nbulk}
\end{equation}
where $N_\text{surf}$ and $N_\text{bulk}$ represent the average number of surface and bulk molecules, respectively, $N = N_\text{bulk} + N_\text{surf}$ is the total number of molecules, and $\omega = - \gamma B$ is the proton Larmor pulsation frequency, where $\gamma/ 2 \pi = 42.58$\,MHz/T is the gyro-magnetic ratio for $^1$H with spin $I = 1/2$, and $B$ the applied magnetic field. 
\Cref{eq:Ntot_Nsurf_Nbulk} is valid in the fast diffusion limit $4 D / H \gg \delta {\cal R}_{1, \text{surf}}$ (where $D$ is the molecular diffusion coefficient and $\delta$ the surface layer thickness), which
is well applicable to the investigated system, as $4 D / H \approx 2$\,m/s using the bulk diffusion coefficient for liquid water at ambient temperature $D \approx 2 \cdot 10^{-9}$\,m$^2$/s and a pore size $H\approx 4$\,nm.
This is significantly larger than $\delta {\cal R}_{1, \text{surf}} \approx 3 \cdot 10^{-10}$\,m/s, where $R_{1, \text{surf}} \approx 1$\,s$^{-1}$ and $\delta \approx 3$\,\AA{} are typical values for liquid water under ambient conditions as will be discussed below.
Importantly, this implies that the fast diffusion limit holds for nanoconfined water, where $D$ and $R_{1, \text{surf}}$ can differ substantially from their values in bulk \cite{faux13a, tsimpanogiannis19a, gravelle23a}.

Levitz proposed to relate the frequency dependence of $R_{1, \text{surf}} (\omega)$ to the intermittent dynamics of the molecules near an interface \cite{levitz05a, levitz13a, levitz19a}: an indicator function $I_{\text{B}, i} (t)$ is first defined for every molecule $i$.
Here, $I_{\text{B}, i} (t)$ equals one when the molecule is adsorbed at the bottom (B) surface of the pore, and zero otherwise (either when the molecule is freely diffusing in the bulk or adsorbed at the top (T) surface).
The single-molecule time autocorrelation function is then calculated as $C_\text{surf} (t) = \sum_{i = 0}^{N} \left< I_{\text{B}, i} (t) I_{\text{B}, i} (0) \right> / \left< I_{\text{B}, i} (t) \right> / N$, where $\left< I_\text{B} (t) \right>$ denotes the probability for the molecule to be adsorbed at the bottom wall with $C_\text{surf} (0) = 1$ and $C_\text{surf} (+ \infty) = N_\text{surf} / N$.
In the fast exchange limit, the spin relaxation at the surface happens predominantly by excursions into the bulk, thus $R_{1, \text{surf}} (\omega)$ is related to the surface spectral density $J_\text{surf} (\omega)$, obtained from the Fourier transform of $C_\text{surf} (t)$.
For dipolar spin-lattice coupling, this yields \cite{abragam61a}
\begin{equation}
R_{1, \text{surf}} (\omega) \propto J_\text{surf} (\omega) + 4 J_\text{surf}  (2 \omega).
\label{eq:RI}
\end{equation}


In order to assess the applicability of the intermittent dynamics approach, we perform atomistic MD simulations of water confined within a slit pore of varying hydrophilicity using the GROMACS simulation package \cite{abraham15a}.
The pore is composed of Lennard-Jones particles placed on a FCC lattice with adjustable surface-water energy $\epsilon_\text{sl}$ (Fig.\,\ref{fig:system}a) and for water the TIP4P/$\epsilon$ model was used \cite{fuentesazcatl14a}, for
further simulation details see \citenum{SI}.
This setup allows for studying a highly hydrophilic surface with $\theta \approx 3^\circ$, where density profiles normal to the solid surface show strong spatial oscillations, indicating a high degree of water structuring (Fig.\,\ref{fig:system}c), as well as a hydrophobic surface with $\theta \approx 127^\circ$, where the density oscillations are less pronounced (Fig.\,\ref{fig:system}d).
Analysis of the average water orientation shows that the molecules in the first layer next to the solid surface tend to align their dipoles parallel to the wall (Fig.\,\ref{fig:S-orientation}), which is expected for interfaces that do not form hydrogen bonds with water \cite{wilson87a, lee84a, ho14a, gravelle22a}.
For details on the contact angles and density and orientation profiles, see \citenum{SI}.

\begin{figure}
\centering
\includegraphics[width=\linewidth]{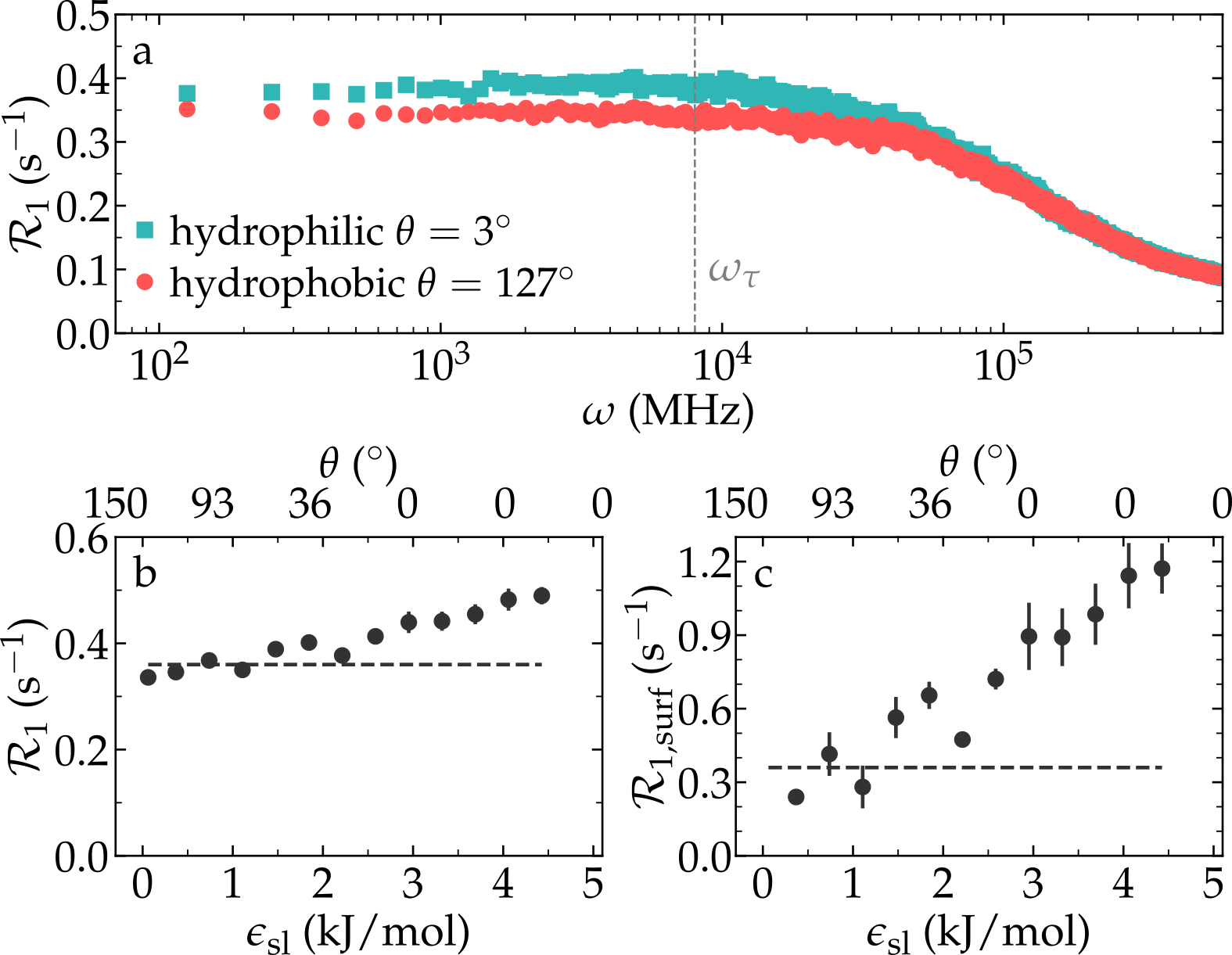}
\caption{
a)~NMR relaxation spectrum ${\cal R}_1$, calculated from fluctuating magnetic dipole-dipole interactions for varying degrees of surface hydrophilicity. The dashed vertical line marks the cutoff frequency $\omega_\tau \approx 8 \cdot 10^3$\,MHz.
b)~NMR relaxation rate ${\cal R}_1$ at a frequency of $\omega = 100$\,MHz as a function of the surface energy $\epsilon_\text{sl}$. 
  The dashed line indicates the bulk water relaxtion time, ${\cal R}_{1, \text{bulk}} = 0.36$\,s$^{-1}$.
  Error bars represent the standard deviation across five independent simulations.
c)~Surface NMR relaxation rate ${\cal R}_{1, \text{surf}}$, calculated using Eq.\,\eqref{eq:Ntot_Nsurf_Nbulk}.
  The corresponding contact angles $\theta$ are shown on the top axis of (b) and (c).
}
\label{fig:NMR}
\end{figure}

The exact $^1$H-NMR relaxation rate, ${\cal R}_1$, for water was calculated from the autocorrelation function of the magnetic dipole-dipole interactions [Eq.\,\eqref{eq:T1RT}] \cite{abragam61a, singer17a,SI}.
The proton relaxation spectra exhibit plateaus at frequencies below $\omega_\tau \approx 8 \cdot 10^3$\,MHz, corresponding to the water rotational and translational timescales \cite{SI}, and rapidly decrease as $\omega$ increases beyond this value (Fig.\,\ref{fig:NMR}a). 
The relaxation rate at the lowest frequency accessible in our simulations, $\omega = 100$\, MHz, corresponds to the typical upper limit of experimental fast field cycling measurements \cite{steele16a} and is shown in Fig.\,\ref{fig:NMR}b. At this frequency, ${\cal R}_1$ increases by about $40\,\%$ with surface hydrophilicity, in line with the expected slowing down of the molecular dynamics near hydrophilic interfaces, where stronger interactions with the surface restrict molecular motion \cite{gravelle23a}.

Using Eq.\,\eqref{eq:Ntot_Nsurf_Nbulk}, the surface contribution ${\cal R}_{1, \text{surf}}$ can be estimated by assuming 
(1) that molecules within the surface layer of width $\delta$ can be identified and 
(2) that molecules outside this layer contribute to ${\cal R}_1$ as if they were in bulk, i.e., with a rate equal to ${\cal R}_{1, \text{bulk}}$.
In detail, to determine $\delta$, the position $z_\rho$ of the first minimum of the density profile was determined for each surface energy $\epsilon_\text{sl}$ (except for $\epsilon_\text{sl} = 0.1$\,kJ/mol where this decomposition is hindered by insufficient fluid structuring), see vertical dashed lines in Fig.\,\ref{fig:system}c,d.
As shown in Fig.\,\ref{fig:NMR}c, ${\cal R}_{1, \text{surf}}$ increases quasi-linearly for increasing surface energy $\epsilon_\text{sl}$, from roughly $0.3\,\text{s}^{-1}$ to about $1.2\,\text{s}^{-1}$.
Importantly, linear extrapolation to the water-vapor interface, $\epsilon_\text{sl} = 0$\,kJ/mol, reveals faster relaxation dynamics for water molecules at such hydrophobic interfaces than in bulk.
Or, vice versa, molecules near a hydrophobic surface contribute less to the relaxation than bulk molecules.
This measured faster dynamics for water confined within a hydrophobic nanopore is also reflected in the corresponding self-diffusion coefficients (Fig.\,\ref{fig:S-MSD}).
Notably, this infers that for the most hydrophobic pores studied, ${\cal R}_1$ shown in Fig.\,\ref{fig:NMR}b is smaller than the value measured for a bulk system in the absence of any surfaces, ${\cal R}_{1, \text{bulk}} = 0.36$\,s$^{-1}$,
aligning with the lower water density and faster molecular motion compared to hydrophilic walls.

\begin{figure}
  \centering
  \includegraphics[width=\linewidth]{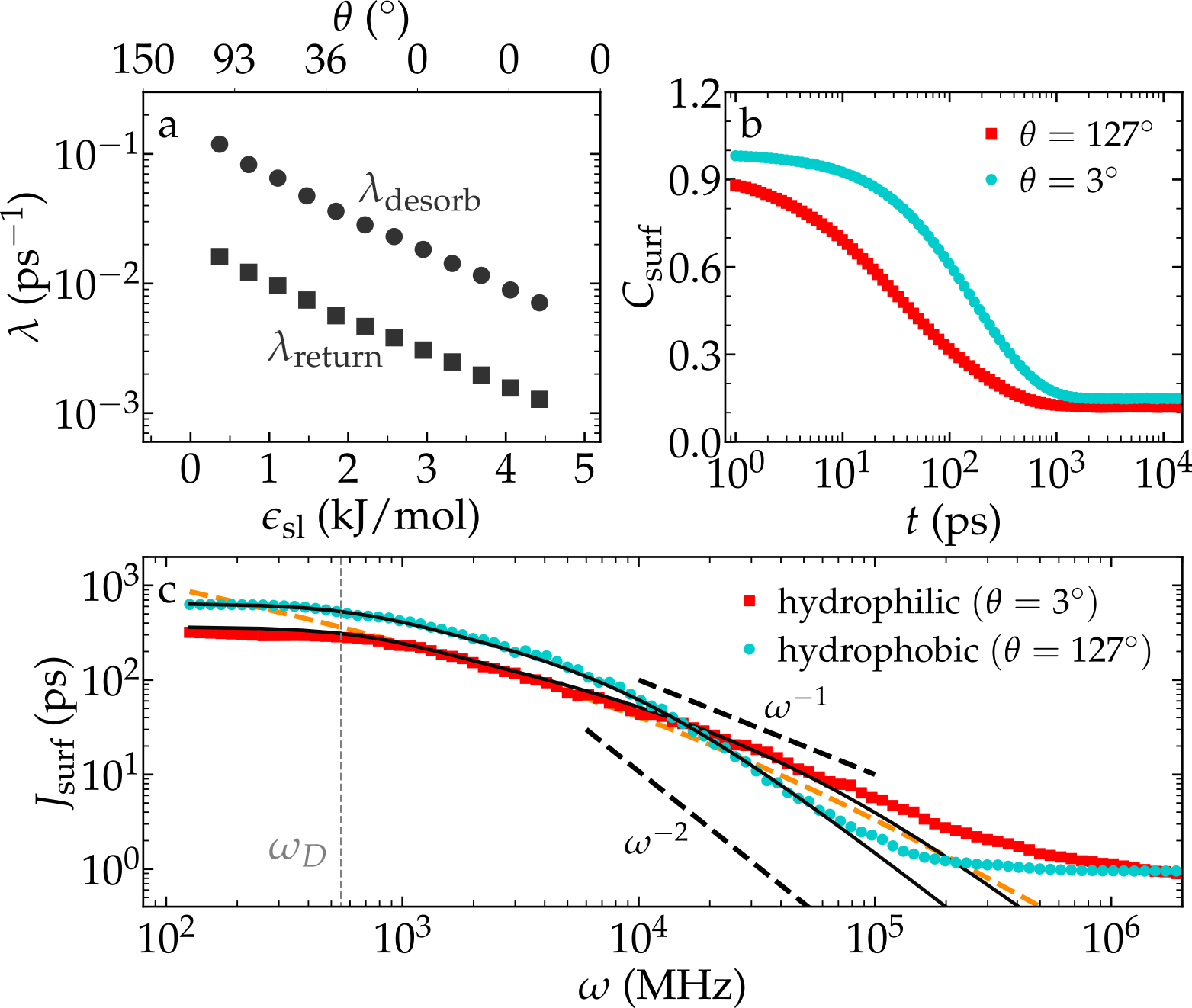}
  \caption{
  a)~Average desorption rate $\lambda_\text{desorb}$ (disks) and return rate $\lambda_\text{return}$ (squares) for varying surface energy $\epsilon_\text{sl}$. 
  b)~Normalized surface correlation functions $C_\text{surf}$ for hydrophobic ($\epsilon_\text{sl} = 0.4$\,kJ/mol) and hydrophilic ($\epsilon_\text{sl} = 3$\,kJ/mol) pores.
  c)~Surface spectra corresponding to the correlation functions shown in panel~b. 
     Symbols represent MD simulation results and the solid dark lines follow from Eq.\,\ref{eq:C_BB-main},
     whereas the orange dashed illustrates Eq.\,\eqref{eq:Jsurf_levitz} for $\omega_\delta = 665$\,GHz corresponding to the data for $\epsilon_\text{sl} = 0.4$\,kJ/mol.
     The dashed vertical line indicates the cutoff frequency, $\omega_D \approx 2 \pi D/H^2$. 
  }
  \label{fig:nmr-intermittent}
  \end{figure}

The exchange of molecules between bulk and surface populations can be assessed in terms of the indicator function $I_{i, \text{B}} (t)$ (Fig.\,\ref{fig:system}b),
which directly allows to determine the average return rate $\lambda_\text{return}$ between two desorptions from the bottom surface and the average desorption rate $\lambda_\text{desorb}$.
As shown in Fig.\,\ref{fig:nmr-intermittent}a, both $\lambda_\text{desorb}$ and $\lambda_\text{return}$ decrease exponentially as $\epsilon_\text{sl}$ increases, 
with the desorption rate being significantly larger than the return rate for the studied system.
Generally, $\lambda_\text{return}$ is expected to depend on pore geometry, size, and the diffusion coefficient of the molecules \cite{levitz19a}. 
Since molecules can adsorb at the top surface before returning to the bottom surface, and because due to mass conservation re-adsorption at the bottom surface is only possible if an adsorbed molecule desorbs simultaneously, $\lambda_\text{return}$ is expected to be proportional to $\lambda_\text{desorb}$, explaining its dependence on surface energy $\epsilon_\text{sl}$ (Fig.\,\ref{fig:nmr-intermittent}a).
The corresponding correlation functions, $C_\text{surf}$, shown in Fig.\,\ref{fig:nmr-intermittent}b decay faster for hydrophobic surfaces compared to hydrophilic ones related to higher desorption and return rates, and thus indicating faster exchange dynamics. 
Although $C_\text{surf}$ is sensitive to the surface population definition, employing different approaches does not significantly affect the present discussion (Figs.\,\ref{fig:S-sensitivity-1}-\ref{fig:S-sensitivity-3}).

The surface spectrum $J_\text{surf}$ shown in Fig.\,\ref{fig:nmr-intermittent}c obtained from Fourier transforming $C_\text{surf}$ reveals three distinct regimes.
For frequencies lower than $\omega_D \approx 500$\,MHz, $J_\text{surf}$ exhibits a plateau that depends on the surface hydrophilicity, corresponding to the limited maximum return time in a finite system \cite{levitz13a}.
Similar low-frequency plateaus for have been observed for non-interacting Brownian particles reversibly adsorbing at the surface of closed nanopores, for which the value of the cutoff was found to be well approximated by the characteristic diffusion frequency $\omega_D \approx 2 \pi D/H^2$ \cite{gravelle19a}. 
Using $\omega_D = 500$\,MHz and $H = 4$\,nm, the diffusion normal to the surface can be estimated as $D = \omega_D H^2 / 2 \pi = 1.3 \cdot 10^{-9}$\,m$^2$/s, in reasonable agreement with the bulk value $D \approx 2.2 \cdot 10^{-9}$\,m$^2$/s for our water model \cite{fuentesazcatl14a}, thus confirming that the value of $\omega_D$ is indeed governed by confinement effects.
For frequencies $\omega$ in between $\omega_D$ and $2 \cdot 10^5]$~MHz, an intermediate regime with $J_\text{surf} \approx \omega^{-\alpha}$ is observed, where $\alpha$ depends on hydrophilicity: 
$\alpha \approx 1$ for the most hydrophobic surface, whereas $\alpha \approx 2$ for the most hydrophilic surface.
This variation is attributed to the interplay between diffusive returns and desorption events, consistent with previous observations in simple pores \cite{gravelle19a}.
For frequencies larger than $\approx 2 \cdot 10^5$\,MHz, a second plateau appears regardless of the surface energy that is related to re-crossing events not captured within the intermittent molecular dynamics description. 

For an unbound surface, the surface spectrum reads \cite{levitz05a}
\begin{equation}
J_\text{surf}^{-1} (\omega) \propto \sqrt{\frac{\omega}{\omega_\delta}} + \frac{\omega}{\omega_\delta} + \frac{1}{2} \left(  \frac{\omega}{\omega_\delta} \right)^{3/2},
\label{eq:Jsurf_levitz}
\end{equation}
where $\omega_\delta = \delta^2 \lambda_\text{desorb}^2 / (2 D)$.
Although this expression links $J_\text{surf}$ to key microscopic parameters governing the intermittent dynamics, it only accounts for single surface (i.e., the limit $H \to \infty$).
As expected for a confined fluid, Eq.\,\eqref{eq:Jsurf_levitz} fails to predict the surface spectrum $J_\text{surf}$ for all frequencies (Fig.\,\ref{fig:nmr-intermittent}\,c), 
especially contrasting the low-frequency plateaus below the diffusive frequency cutoff, where Eq.\,\eqref{eq:Jsurf_levitz} predicts $J_\text{surf} \sim 1/\sqrt{\omega}$ as $\omega \to 0$. We here derive an alternative formalism for $J_\text{surf}$ based on the first return passage time of particles between successive adsorption and desorption events in a confined space \cite{gravelle19a}. 
This involves counting all possible desorption and re-adsorption events that a molecule undergoes over time, yielding the Laplace transform of the correlation function \cite{gravelle19a}
\begin{equation}
\tilde{C}_\text{B} (s)
= \dfrac{\lambda_\text{desorb}^{-1} \tilde{\psi}_\text{B} (s) \tilde{J}_\text{B}^* (s)}
{1 - \tilde{\psi}_\text{B} (s) \tilde{J}_\text{B}^* (s) \tilde{J}_\text{B $\to$ T} (s) \tilde{J}_\text{T}^* (s) \tilde{\psi}_\text{T} (s) \tilde{J}_\text{T $\to$ B} (s)},
\label{eq:C_BB-main}
\end{equation}
where $\tilde{J}_\text{X}^* (s) = [1 - \tilde{\psi}_\text{X} (s)  \tilde{J}_\text{X $\to$ Y} (s)]^{-1}$ with X,Y = T or B. 
$\tilde{\psi}_\text{X} (s) = \lambda_\text{desorb} / (\lambda_\text{desorb} + s)$ are the Laplace transforms of the survival probability functions $\psi_\text{X} (t)$,
which describe the time until a molecule adsorbed at wall X desorbs and reveal an extended exponential tail (Fig.\,\ref{fig:S-probability}) \cite{levitz13a}.
The functions $\tilde{J}_\text{X $\to$ Y} (s)$ are the first-return distributions for a molecule leaving the surface X before adsorbing to Y (where X and Y can be the same or different surfaces). The equations for $\tilde{J}_\text{X $\to$ Y} (s)$ are derived by solving the 1D diffusion equation for the Greens function, $\partial_t {G} (z, t | z_0) = D \partial_z^2 {G} (z, t | z_0)$ in Laplace space,
\begin{equation}
(s - D \partial_z^2) {\tilde G} (z, s | z_0) = \delta (z - z_0),
\label{eq:diffusion-laplace-main}
\end{equation}
with the  initial condition ${\tilde G} (z, t=0 | z_0) = \delta (z - z_0)$
and reactive boundary conditions at the walls.
For the bottom wall at $z=0$, the latter is $D \partial_z \tilde{G} (0, s | z_0) = k \tilde{G} (0, s | z_0)$ with
\begin{equation}
k = \dfrac{ N_\text{surf}}{N_\text{bulk}} H \lambda_\text{desorb},
\label{eq:k-main}
\end{equation}
which decreases from $\approx 30$\,m/s for the most hydrophobic surface considered here, to $\approx 2$\,m/s for the most hydrophilic (Fig.\,\ref{fig:S-k}).
Full calculations for the first return passage time statistics $\tilde{J}_\text{X $\to$ Y} (s)$ are given in \cite{SI}.
The surface spectra, following as $J_\text{surf} (\omega) = \tilde C_\text{B} (i \omega) + \tilde C_\text{B} (- i \omega)$, are shown as solid lines in Fig.\,\ref{fig:nmr-intermittent}\,c,
revealing excellent agreement of the first passage calculations in Eqs.\,(\ref{eq:C_BB-main}-\ref{eq:k-main}) with the MD results for both hydrophilic and hydrophobic surfaces at frequencies $\omega < 5 \cdot 10^4$\,MHz. 
The saturation of $J_\text{surf}$ at higher frequency, attributed to re-crossing, is not captured within this model.

\begin{figure}
\centering
\includegraphics[width=\linewidth]{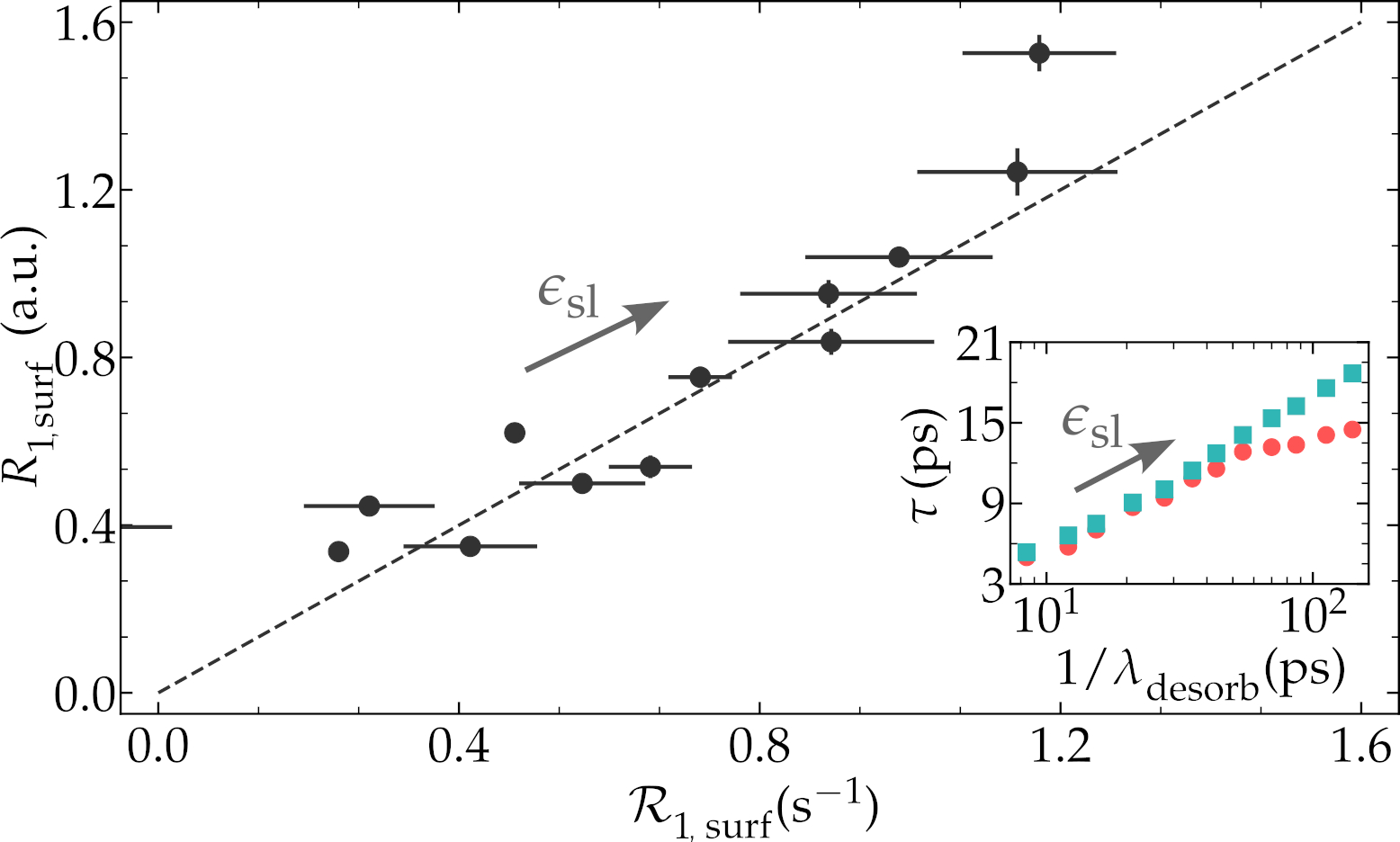}
\caption{Comparison between $R_{1, \text{surf}}$, as calculated using the intermittent model [Eqs.\,(\ref{eq:RI}, \ref{eq:C_BB-main}-\ref{eq:k-main})], and ${\cal R}_{1, \text{surf}}$ from dipole-dipole correlations [Eq.\,\eqref{eq:T1RT}] at frequency $\omega = 100$\,MHz. Error bars indicate the standard deviation from five statistically independent simulation realizations. Inset: Average rotational (red disks) and translational (cyan squares) times, $\tau_\text{rot}$ and $\tau_\text{trans}$ [Eqs.\,(\ref{eq:tau_intra}-\ref{eq:tau_inter})], respectively, of water molecules as a function of the desorption time $1/\lambda_\text{desorb}$.}
\label{fig:nmr-final-comparison}
\end{figure}

Having assessed the validity of the first passage time approach for surface spectra,
we now turn to the applicability of intermittent molecular dynamics for understanding and predicting NMR relaxation rates in confined systems.
In detail, we compared the evolution of $R_{1, \text{surf}}$ with the surface interaction strength $\epsilon_\text{sl}$ determined from [Eq.\,\eqref{eq:RI}]
with ${\cal R}_{1, \text{surf}}$ derived from magnetic dipole-dipole interactions [see Eq.\,\eqref{eq:T1RT} in \cite{SI} for details].
Our results show that $R_{1, \text{surf}}$ successfully estimates the increase in ${\cal R}_{1, \text{surf}}$ with $\epsilon_\text{sl}$ (Fig.\,\ref{fig:nmr-final-comparison}).
The main discrepancy between the two approaches appears at the highest surface energy $\epsilon_\text{sl} = 4.4$\,kJ/mol, where $R_{1, \text{surf}}$ 
is overestimated by about $15\,\%$, which we attribute to the long tail exponential survival probability for strong surface interactions (Fig.\,\ref{fig:S-probability}).
Quantifying the NMR relaxation rate in terms of intermittent motion reveals intriguing accuracy and allows for predictions with changing surface interaction or pore size.
This is at first sight surprising since $R_{1, \text{surf}}$ derived from the surface spectra shows variation with $\omega$ down to a cutoff frequency set by slow diffusion, $\omega_D \approx 500$\,MHz (Fig.\,\ref{fig:nmr-intermittent}c), consistent with the expected sensitivity of $R_{1, \text{surf}}$ to molecular desorption and re-adsorption events.
In contrast, ${\cal R}_1$ varies only down to $\omega_\tau \approx 8 \cdot 10^3\,\text{MHz} \gg \omega_D$ (Fig.\,\ref{fig:NMR}a),
close to the frequency of molecular motion of water at ambient conditions (where the rotational time $\tau_\text{rot} = 2.7$\,ps and the translational time $\tau_\text{trans} = 4$\,ps \cite{singer17a, SI}).
It is tempting to explain that the intermittent model successfully captures the trend for $R_{1, \text{surf}}$ by its natural incorporation of the desorption frequency $\lambda_\text{desorb}$, which includes the confined molecular motions.
Notably, $\lambda_\text{desorb}$ varies alongside $\tau_\text{rot}$ and $\tau_\text{trans}$ as $\epsilon_\text{sl}$ increases, supporting this connection between molecular confinement and relaxation rate (Inset in Fig.\,\ref{fig:nmr-final-comparison}) \cite{SI}.

Summarizing, the behavior of the relaxation rate ${\cal R}1$ can be predicted based on the intermittent dynamics of the reversibly adsorbing molecules, linking microscopic parameters such as pore size $H$ and desorption frequency $\lambda\text{desorb}$ to NMR relaxation rates.
This relationship is probed through MD simulations of water confined within planar nanopores with adjustable hydrophilicity. The surface spectrum, $J_\text{surf}$, can be predicted using the first return passage time of molecules between successive adsorption and desorption events.
Since the first passage time calculations can be adapted to any closed and open geometries, as well as to arbitrary molecular interactions, these findings support the integration of experimental and theoretical NMR approaches for studying fluids at interfaces. Additionally, the intermittent model could be extended to more complex systems, such as interfaces with ill-defined surfaces \cite{roosen16a} or pores with complex topologies \cite{bousige21a},  provided that the surface populations can be detected. 
This detection can be generalized, for instance, using ITIM \cite{partay08c} 
(see Fig.\,\ref{fig:PITIM} for a proof of concept).

\begin{acknowledgments}
We thank the Deutsche Forschungsgemeinschaft (DFG, German Research 
Foundation) for funding via project Number 327154368 - SFB 1313. A.S. and C.H. acknowledge 
funding from the DFG under Germany's Excellence Strategy-EXC 2075-390740016 and 
support by the Stuttgart Center for Simulation Science (SimTech). S.G. and B.C. acknowledge
funding from the European Union's Horizon 2020 research and innovation programme
under the Marie Skłodowska-Curie grant agreement N$^\circ\;101065060$.
\end{acknowledgments}

\section*{Data Availability Statement}

\noindent GROMACS input files, Python scripts for generating the initial configuration and force field parameters are openly available from the DaRUS repository \cite{darus23a}.

\nocite{sega18a, bussi07a, berendsen84a, essmann95a, hess97a, shi09a, michaudagrawal11a, maicos, cowan97a, bloembergen48a, torrey53a, grivet05a, becher21a}

\bibliography{icp,data}
\renewcommand{\bibname}{} 

\clearpage

\onecolumngrid

\renewcommand{\thefigure}{S\arabic{figure}}
\renewcommand{\theequation}{S\arabic{equation}}
\setcounter{figure}{0}
\setcounter{equation}{0}

\begin{figure*}[h!]
\includegraphics[width=0.45\linewidth]{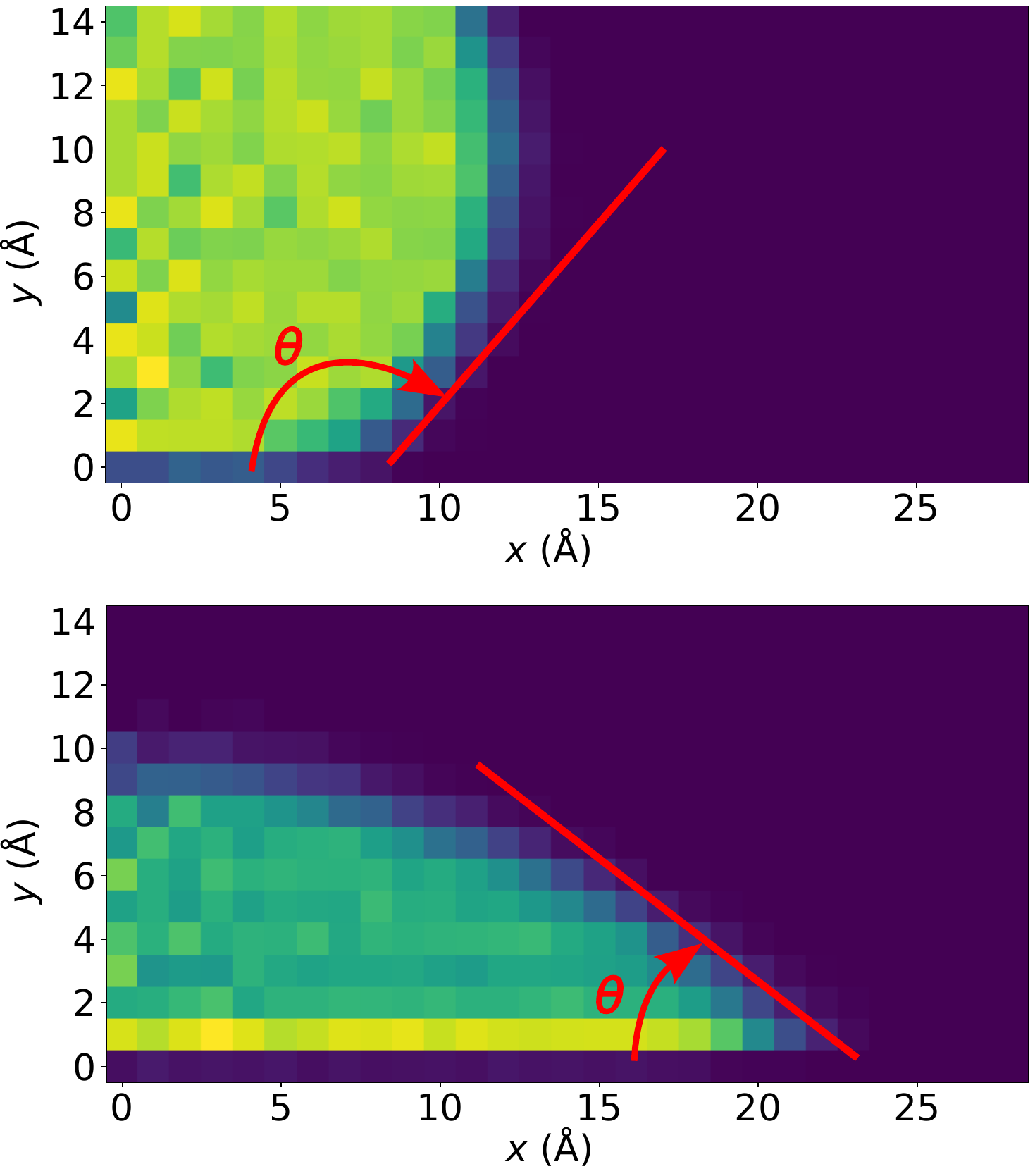}
\caption{Density profiles of water droplets used for contact angle measurements, with yellow indicating high density and blue indicating zero density. The solid surface is located at $y<0$. Profiles are shown for two solid-liquid energies: $\epsilon_\text{sl} = 0.4$\,kJ/mol (top) and $\epsilon_\text{sl} = 2$\,kJ/mol (bottom).}
\label{fig:S-contact-angle}
\end{figure*}

\begin{figure*}[h!]
\includegraphics[width=0.5\linewidth]{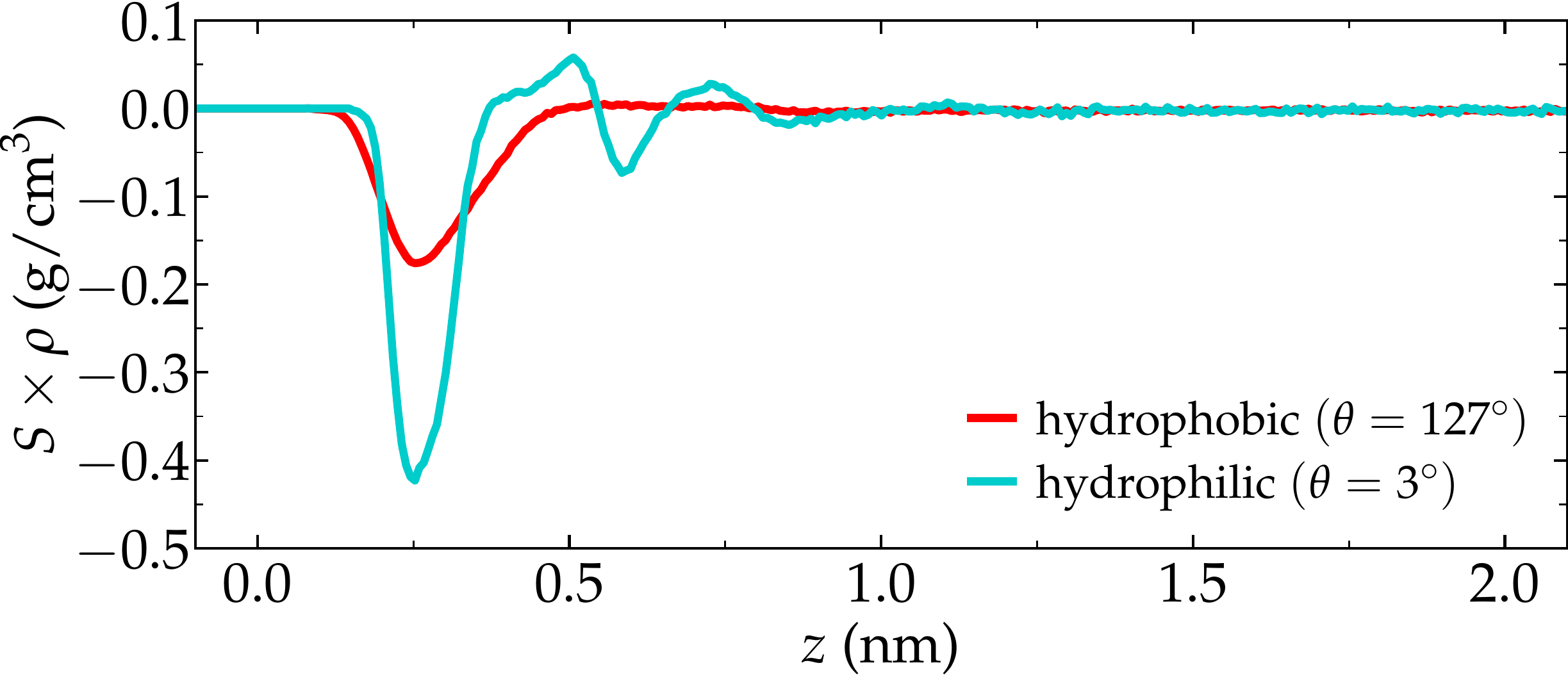}
\caption{
Orientation parameter $S$ [Eq.\,\eqref{eq:S}] multiplied by the water density $\rho$ as a function of $z$, shown for a hydrophilic surface with $\epsilon_\text{sl} = 2.6$\,kJ/mol ($\theta = 3^\circ$) and hydrophobic surface with $\epsilon_\text{sl} = 0.4$\,kJ/mol ($\theta = 127^\circ$). Here, $z=0$ corresponds to the position of the first atomic layer from the wall.
}
\label{fig:S-orientation}
\end{figure*}

\begin{figure*}
\centering
\includegraphics[width=0.55\linewidth]{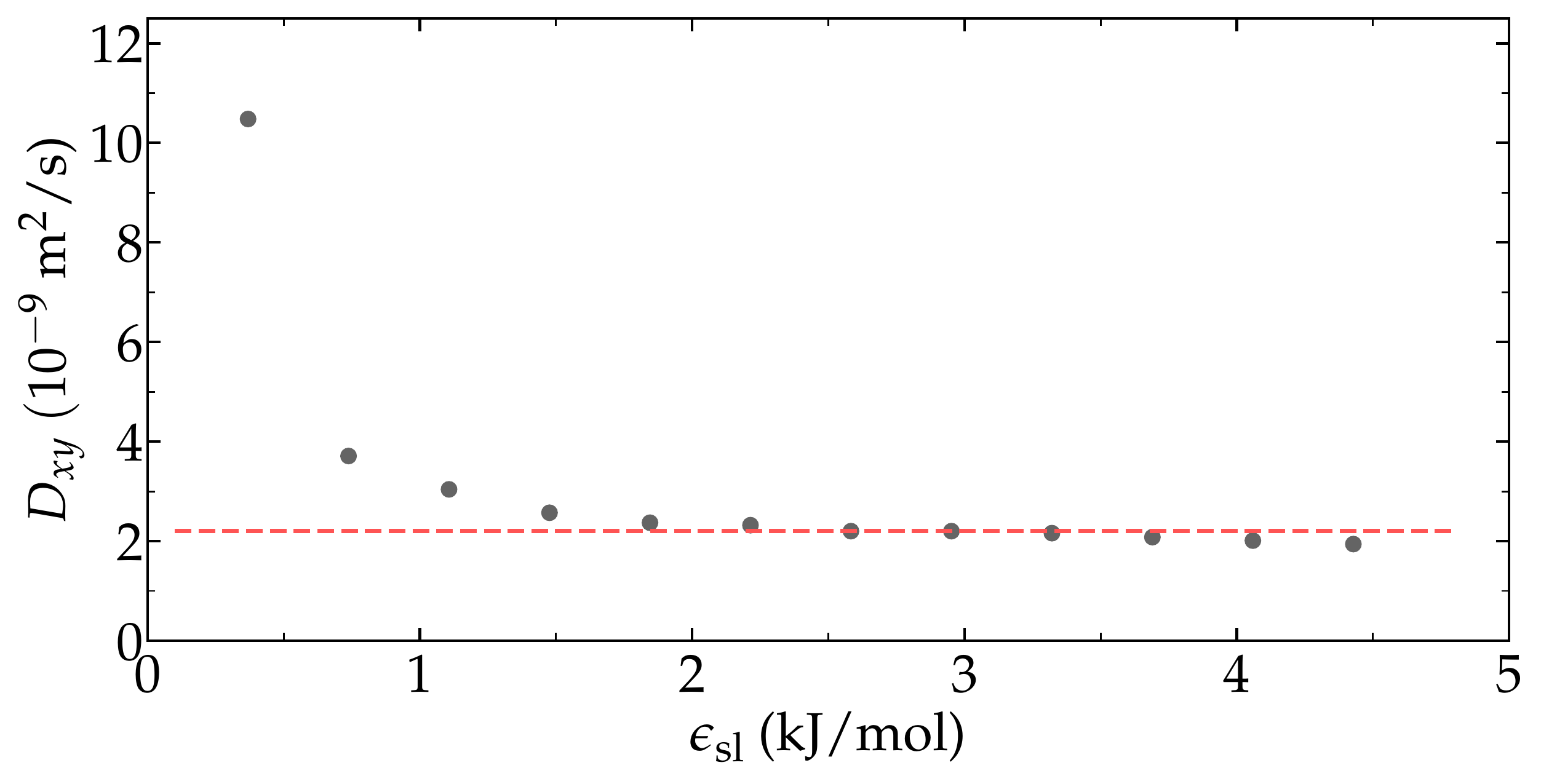}
\caption{Self-diffusion coefficients $D_{xy}$ in the x-y direction as a function of the surface energy $\epsilon_\text{sl}$. $D_{xy}$ was obtained for water by evaluating the lateral mean square displacement, $\text{MSD} (t) = \left< | \mathbf{r}_{xy}(t) - \mathbf{r}_{xy}(0) |^2 \right>$, where $\mathbf{r}_{xy} = (x, y)$. The diffusion coefficient was calculated as $D_{xy} = \left< | \mathbf{r}_{xy}(t) - \mathbf{r}_{xy}(0) |^2 \right> / 4 t$ in the long time limit $t \to \infty$. The horizontal dashed line is the reference bulk value $D \approx 2.2 \cdot 10^{-9}$\,m$^2$/s for our water model \cite{fuentesazcatl14a}.
}
\label{fig:S-MSD}
\end{figure*}

\begin{figure*}[h!]
\includegraphics[width=0.55\linewidth]{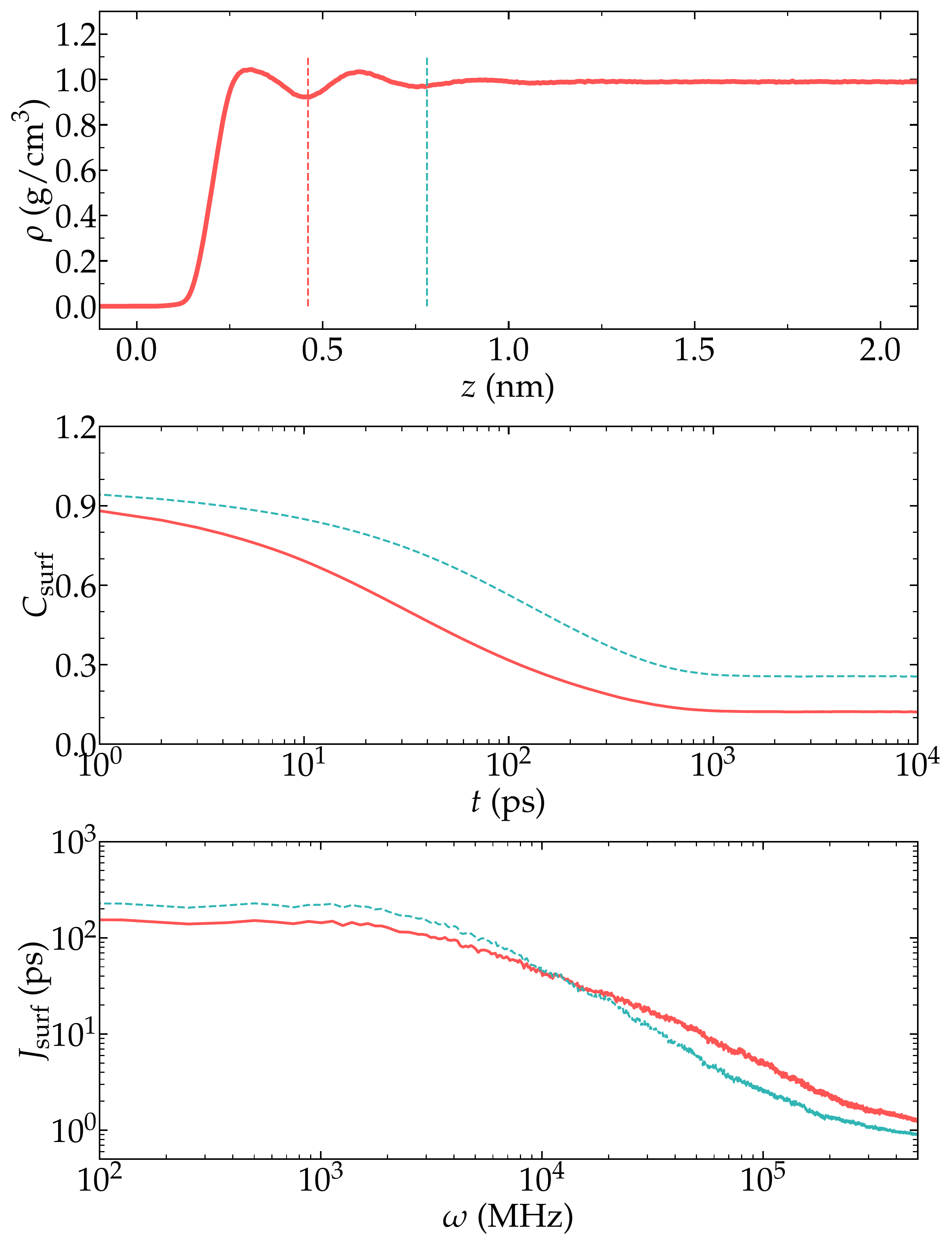}
\caption{Water density profile (top) for $\epsilon_\text{sl} = 0.4$\,kJ/mol. The solid vertical line indicates the dividing line between surface and the bulk populations as defined in the main text, while the dashed line demarcates a surface population made of the first two density layers. Corresponding surface correlation functions, $C_\text{surf}$ (middle), and surface spectra $J_\text{surf}$ (bottom), are shown for the two dividing line positions illustrated in the top panel.}
\label{fig:S-sensitivity-1}
\end{figure*}

\begin{figure*}[h!]
\includegraphics[width=0.55\linewidth]{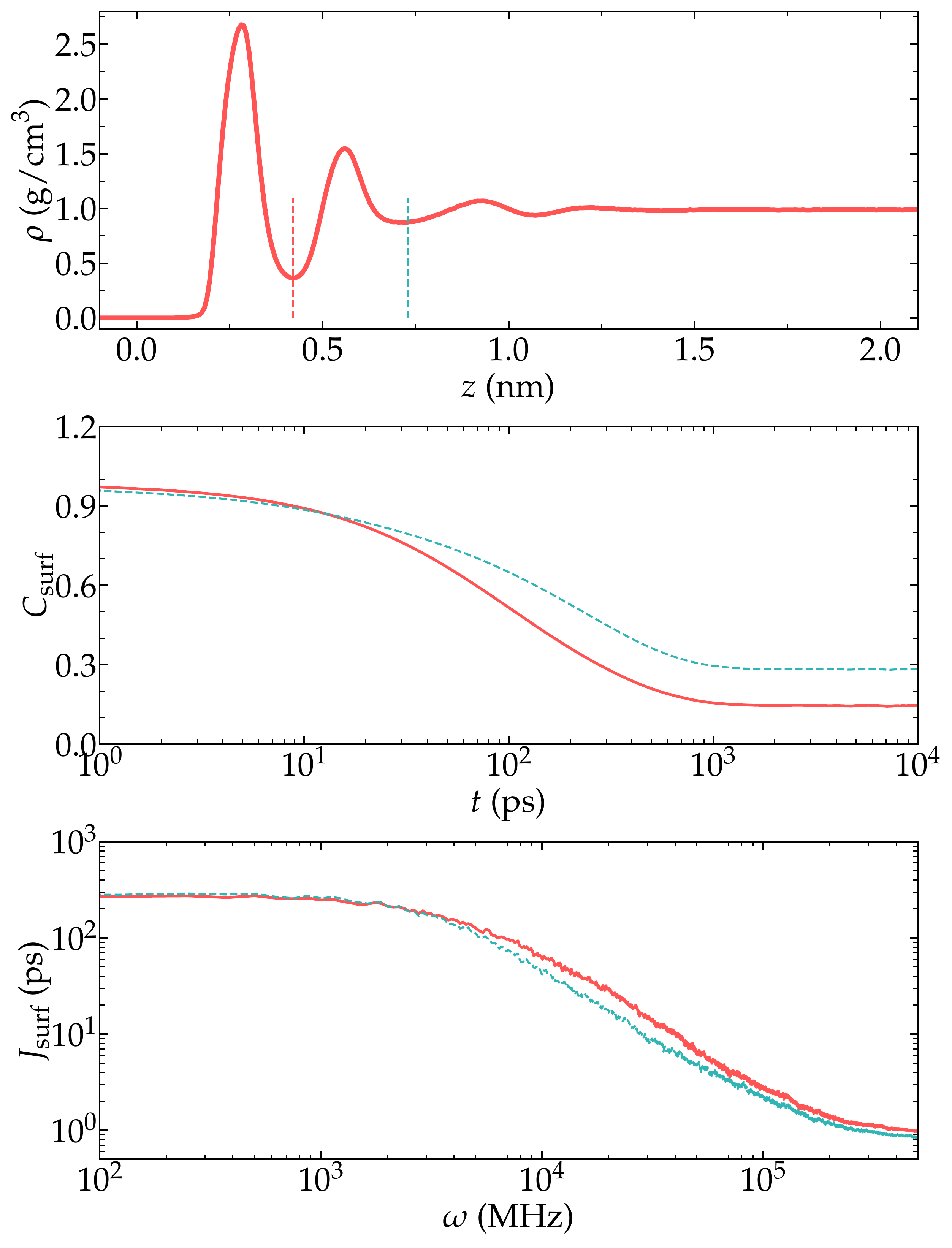}
\caption{Water density profile (top) for $\epsilon_\text{sl} = 2.4$\,kJ/mol. The solid vertical line indicates the dividing line between surface and the bulk populations as defined in the main text, while the dashed line demarcates a surface population made of the first two density layers. Corresponding surface correlation functions, $C_\text{surf}$ (middle), and surface spectra $J_\text{surf}$ (bottom), are shown for the two dividing line positions illustrated in the top panel.}
\label{fig:S-sensitivity-2}
\end{figure*}

\begin{figure*}[h!]
\includegraphics[width=0.55\linewidth]{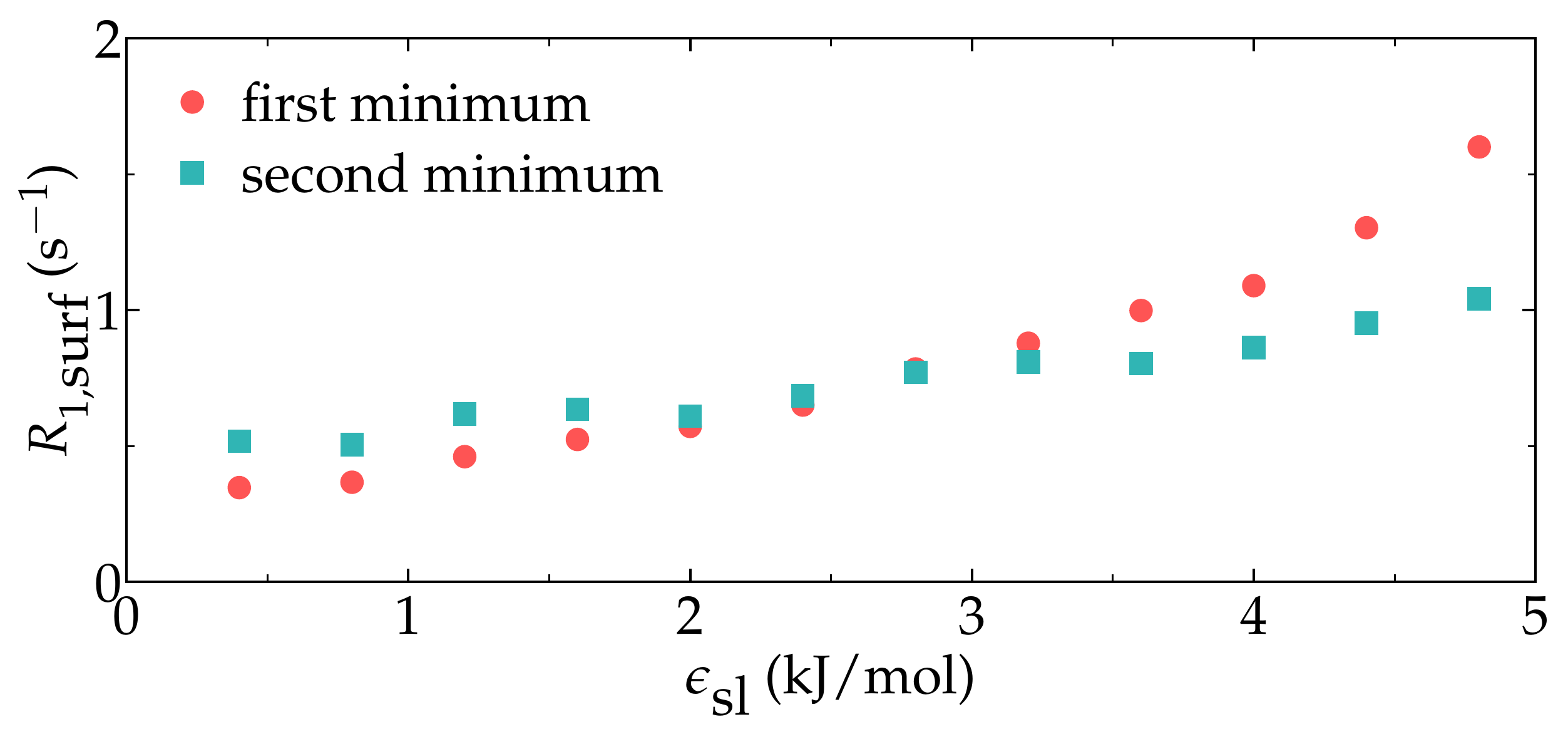}
\caption{Surface NMR relaxation rate $R_{1, \text{surf}}$ as calculated using Eqs.\,(2-4) of the main text, shown as a function of the LJ surface energy, $\epsilon_\text{sl}$, for two positions of the dividing line between the surface and bulk
populations. The `first minimum' data represents a surface population defined by the first density layer, while the `second minimum' data represents a surface population made of the two first density layers.}
\label{fig:S-sensitivity-3}
\end{figure*}

\begin{figure*}[h!]
\includegraphics[width=0.55\linewidth]{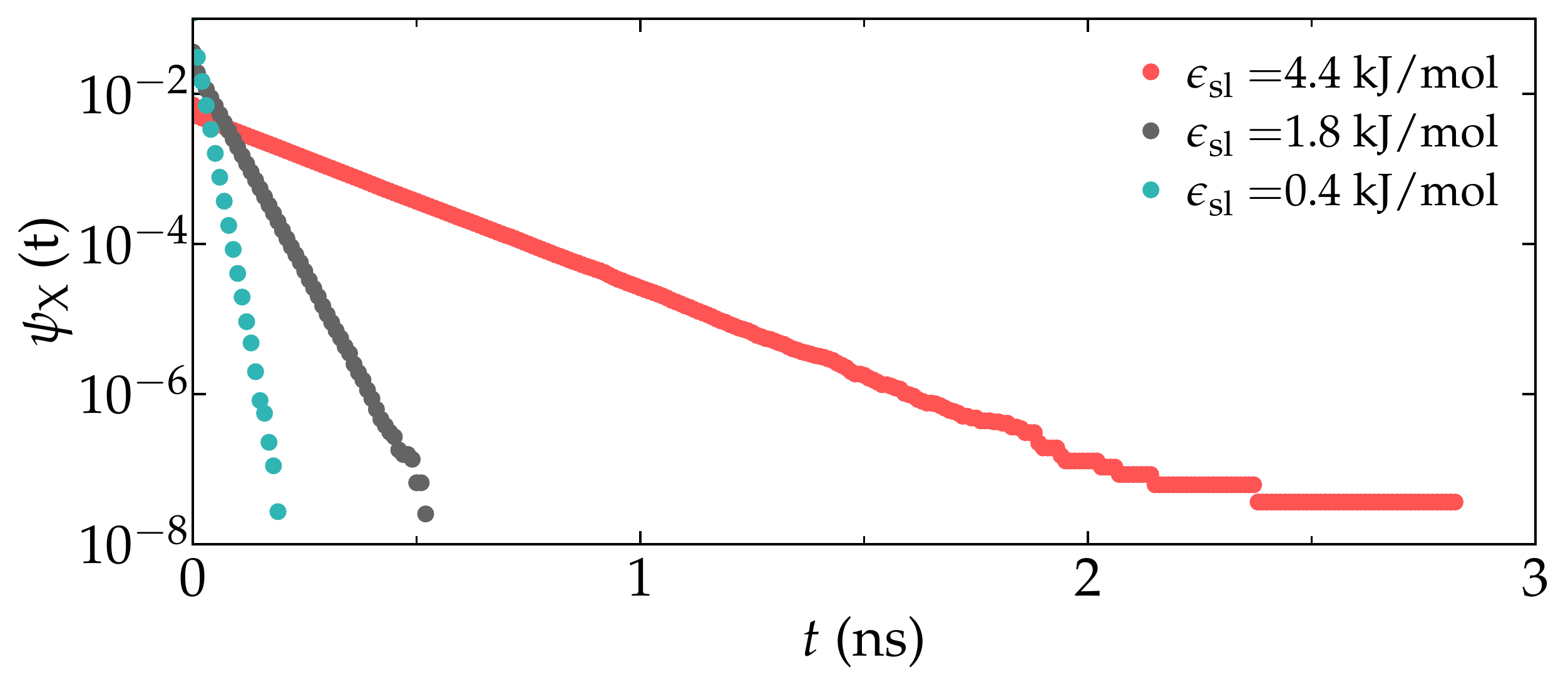}
\caption{Probability distribution $\psi_\text{X} (t)$ for an adsorbed molecule to desorb from a surface X at time $t$, as extracted from MD for varying solid-liquid interaction energies $\epsilon_\text{sl}$ (see the legend).
}
\label{fig:S-probability}
\end{figure*}

\begin{figure*}[h!]
\includegraphics[width=0.55\linewidth]{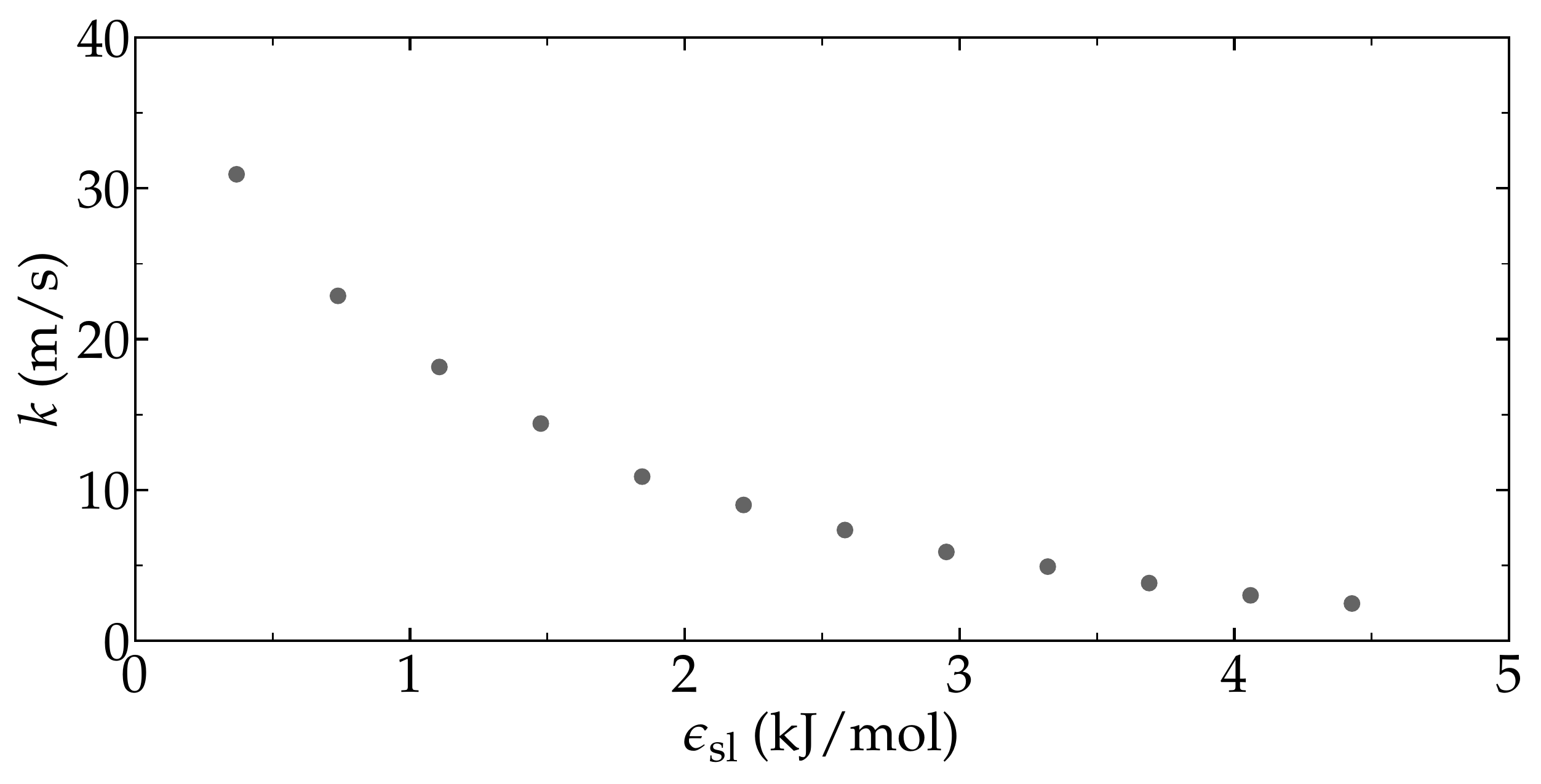}
\caption{Rate constant $k$ as a function of the surface energy $\epsilon_\text{sl}$.}
\label{fig:S-k}
\end{figure*}

\begin{figure*}
\centering
\includegraphics[width=0.8\linewidth]{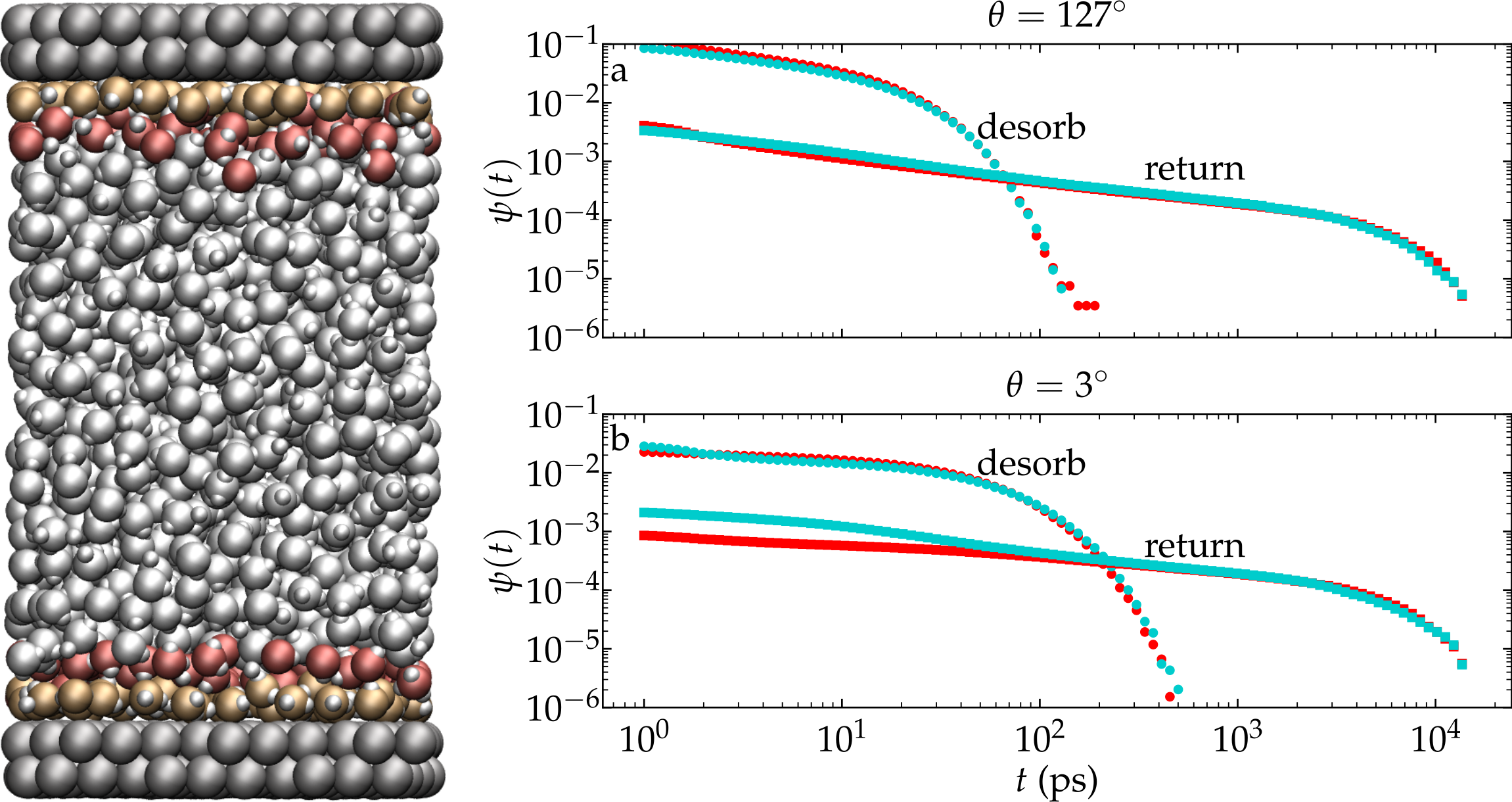}
\caption{Left: Snapshot of the molecular dynamics system with the two first density layers adjacent to the solid surfaces detected using the ITIM algorithm. Right: desorption (disks) and return (squares) distributions $\psi$ when the surface populations are detected using ITIM (cyan) or from the position of the first density depth $z_\rho$ (red) for two values of the contact angle, $\theta = 127^\circ$ (a), and $\theta = 3^\circ$ (b).}
\label{fig:PITIM}
\end{figure*}

\section{Supplemental figures \ref{fig:S-contact-angle}-\ref{fig:PITIM}}

\section{Molecular dynamics simulations}

\noindent Molecular dynamics (MD) simulations of water confined within nanoslit pores (see Fig.\,\ref{fig:system} in the main text) were performed using the GROMACS simulation package \cite{abraham15a}. The lateral dimensions of the system were $\ell_x = 2.8$\,nm and $\ell_y = 2.8$\,nm, and the distance between the two surfaces of the wall was $H \approx 4$\,nm. The solid wall consisted of 560 atoms arranged in a face-centered cubic lattice with parameter $4.04$\,\AA, and had a thickness of approximately $2$\,nm. A total of $N = 1150$ water molecules were placed within the pore. The energy of the LJ potential between the oxygen of the water molecules and the atoms of the wall was varied from $\epsilon_\text{sl} = 0.1$\,kJ/mol (nonwetting with contact angle $\theta \approx 150^\circ$) to $\epsilon_\text{sl} = 4.5$\,kJ/mol (fully wetting liquid with contact angle $\theta = 0^\circ$). The LJ potential width was $\sigma_\text{sl} = 2.9$\,\AA, and the TIP4P/$\epsilon$ model was used for water \cite{fuentesazcatl14a}.

The initial configuration was created with a custom Python script, positioning the water molecules in a square lattice adjacent to the solid surface. To ensure the system was well-equilibrated, the following steps were executed:
\begin{itemize}
\item step 1: The system was first relaxed at temperature $T = 300$\,K during $2.5$\,ps with a timestep of 0.5\,fs. For all the following steps, a temperature of $T = 300$\,K is imposed.
\item step 2: The system was further relaxed during $50$\,ps with a timestep of 1\,fs.
\item step 3: To establish solid-liquid contact and eliminate gas bubbles, the system was compressed using anisotropic pressure coupling with an imposed pressure $P_z = 1000$\,bar along the $z$ axis, an imposed pressure $P_{xy} = 1$\,bar along the $x$ and $y$ axis for a duration of $100$\,ps and with a timestep of 2\,fs. For all the following steps, a timestep of 2\,fs is used. 
\item step 4: The system was then relaxed for 1\,ns with an imposed pressure of $1$\,bar in all three spatial directions.
\item step 5: A final equilibration was conducted over 100\,ps without pressure coupling.
\item step 6: A production run of 50\,ns was performed without pressure coupling, with the configurations saved every 1\,ps.
\end{itemize}
For each case (i.e., each value of $\epsilon_\text{sl}$), steps 1 to 6 were reproduced five times, and results were averaged across these five runs. Error bars represent the standard deviation across these independent simulations. Throughout all six steps, temperature control was implemented using the CSVR thermostat \cite{bussi07a} with a default time constant of $t_\text{CSVR} = 0.5$\,ps. In steps 3 and 4, the pressure was controlled using the Berendsen barostat \cite{berendsen84a} with a time constant of $1$\,ps. Long-range electrostatic interactions were managed using the smooth particle mesh Ewald method (SPME) \cite{essmann95a} and LINCS algorithm with an expansion order of $N_\text{LINCS} = 4$ was used to maintain the geometry of the water molecules \cite{hess97a}.  LJ interactions were truncated at a cutoff of $r_\text{LJ} = 1.2$\,nm, with real-space electrostatic interaction truncated at $r_\text{C} = r_\text{LJ}$. Periodic boundary conditions were employed in all directions. The input scripts can be downloaded from the DaRUS open repository \cite{darus23a}.

\section{Contact angle measurement}

\noindent Contact angle measurements were performed using the same protocol described in Ref.\,\citenum{shi09a}. Water droplets composed of 2000 water molecules were placed on a solid surface with lateral dimensions of $11 \times 11\,\text{nm}^2$. The system was equilibrated at a temperature of 300\,K, and the density was recorded for 1\,ns. The reference position $y = 0$ was calculated from the positions of the topmost atoms of the wall, and the contact angle was estimated by fitting a straight line to the density profile at $y \to 0$. Examples of density profiles obtained for two different values of the solid-liquid energy $\epsilon_\text{sl}$ are shown in Fig.\,\ref{fig:S-contact-angle}. For each value of $\epsilon_\text{sl}$, five statistically independent simulations were performed, and the average contact angle was calculated.

\section{Density profiles}

\noindent The water density profiles in the direction $z$, normal to the solid wall, were extracted using MDAnalysis \cite{michaudagrawal11a} in conjunction with MAICoS Python toolkit \cite{maicos}.

\section{Orientation profiles}

The orientation profile of the water was extracted by defining the orientation parameter $S$ based on the second Legendre polynomial:
\begin{equation}
S (z) = \dfrac{3}{2} \left< \cos ^2 \left(\theta_\text{d} (z)\right) - \dfrac{1}{2} \right>_{\theta_\text{d}},
\label{eq:S}
\end{equation}
where $\theta_\text{d} (z)$ is the angle between the dipole moment of a water molecule and the normal to the solid surface at the position $z$. The bracket $\left< \cdot \right>_{\theta_\text{d}}$ denotes the ensemble average over all angles $\theta_\text{d}$ (see the orientation profiles in Fig.\,\ref{fig:S-orientation}).

\section{Measurement of the molecular times $\tau_\text{intra}$ and $\tau_\text{inter}$}

The autocorrelation functions (Eq.\,\eqref{eq:Gm}) can be split into an intra-molecular and an inter-molecular components \cite{singer17a}:
\begin{equation}
{\cal G}^{(m)}_\text{intra} (t) = \dfrac{1}{N_\text{intra}} \sum_{i \ne j}^{N_\text{intra}} \left< {\cal F}_{ij}^{(m)} (t) {\cal F}_{ij}^{(m)} (0)  \right>,
\label{eq:Gm_intra}
\end{equation}
\begin{equation}
{\cal G}^{(m)}_\text{inter} (t) = \dfrac{1}{N_\text{inter}} \sum_{i \ne j}^{N_\text{inter}} \left< {\cal F}_{ij}^{(m)} (t) {\cal F}_{ij}^{(m)} (0)  \right>,
\label{eq:Gm_inter}
\end{equation}
where $N_\text{inter}$ and $N_\text{intra}$ denote the respective partial ensembles for intramolecular or intermolecular dipole-dipole interactions. The average rotational and translational times are calculated as \cite{cowan97a, singer17a}
\begin{equation}
\tau_\text{intra} = \dfrac{1}{{\cal G}^{(0)}_\text{intra} (0)} \int_0^\infty {\cal G}^{(0)}_\text{intra} (t) \mathrm{d} t,
\label{eq:tau_intra}
\end{equation}
\begin{equation}
\tau_\text{inter} = \dfrac{1}{{\cal G}^{(0)}_\text{inter} (0)} \int_0^\infty {\cal G}^{(0)}_\text{inter} (t) \mathrm{d} t.
\label{eq:tau_inter}
\end{equation}

\section{$^1$H NMR from magnetic dipole-dipole interactions}

\noindent The \textit{exact} $^1$H NMR relaxation rate ${\cal R}_1$ can be calculated from the autocorrelation functions ${\cal G}^{(m)} (t)$ of fluctuating magnetic dipole-dipole interactions \cite{bloembergen48a, torrey53a, abragam61a, cowan97a, grivet05a, singer17a, becher21a},
\begin{equation}
{\cal G}^{(m)} (t) = \dfrac{1}{N} \sum_{i \ne j}^{N} \left< {\cal F}_{ij}^{(m)} (t) {\cal F}_{ij}^{(m)} (0)  \right>.
\label{eq:Gm}
\end{equation}
The ensemble average in Eq.\,\eqref{eq:Gm} is performed by a double summation over spin pair $ij$ with $i \ne j$. ${\cal F}_{ij}^{(m)} (t)$ are functions of the vector $\boldsymbol{r}_{ij}$ between the positions of the spins $i$ and $j$,
\begin{equation}
{\cal F}_{ij}^{(m)} (t) = \alpha_m 
\dfrac{1}{r_{ij}^3 (t)}
{\cal Y}^{(m)}_2 (\theta_{ij} (t), \varphi_{ij} (t)),
\label{eq:Fij}
\end{equation}
where $\theta_{ij} (t)$ and  $\varphi_{ij} (t)$ are respectively the polar and the azimuthal angles with respect to the laboratory axes, assuming that the applied static magnetic field is parallel to $\boldsymbol{e}_z$. $r_{ij} (t)$ is the nuclear spin distance separation.
${\cal Y}^{(m)}_2$ are the normalized spherical harmonics with $\ell = 2$, and
$\alpha_0 = \sqrt{16 \pi /5}$, $\alpha_1 = \sqrt{8 \pi /15}$,
$\alpha_2 = \sqrt{32 \pi /15}.$
The spectral density ${\cal J}^{(m)} (\omega)$ is obtained from the Fourier transform of ${\cal G}^{(m)} (t)$, from which the relaxation rate ${\cal R}_1$ can be calculated as
\begin{eqnarray}
{\cal R}_1 (\omega_0)  = {\cal K} \left[{\cal J}^{(1)} (\omega_0) + {\cal J}^{(2)} (2 \omega_0) \right],
\label{eq:T1RT}
\end{eqnarray}
where
\begin{equation}
{\cal K} = \dfrac{3}{2}\left(\dfrac{\mu_0}{4 \pi}\right)^2 \hbar^2 \gamma^4 I (I+1),
\label{eq:Kb}
\end{equation}
with $\mu_0$ the vacuum permeability, $\hbar$ the reduced Planck constant, 
and $\gamma/ 2 \pi = 42.58$\,MHz/T the gyro-magnetic ratio for $^1$H with spin $I = 1/2$. For bulk systems in absence of interface, ${\cal J}^{(0)} = 6 {\cal J}^{(1)} = 6 {\cal J}^{(2)} / 4$ \cite{becher21a} and Eq.\,\eqref{eq:T1RT} becomes
\begin{equation}
{\cal R}_1 (\omega_0)  = \frac{1}{6} {\cal K} \left[{\cal J}^{(0)} (\omega_0) + 4 {\cal J}^{(0)} (2 \omega_0) \right].
\label{eq:T1RT_bulk}
\end{equation}

\section{NMR from intermittent dynamics}

\noindent Levitz proposed an expression to describe the frequency dependence of the surface NMR relaxation, $R_{1, \text{surf}} (\omega_0)$, based on the intermittent dynamics of the molecules near the interface \cite{levitz05a, levitz13a, levitz19a}. To achieve this, an indicator function $I_{\text{X}, i} (t)$ is defined for each molecule $i$, where $I_{\text{X}, i} (t)$  equals one when the molecule is adsorbed at the surface X of interest, and zero when the molecule is freely diffusing in the bulk or adsorbed at another surface. The time auto-correlation function for all $I_{\text{X}, i} (t)$ functions is then computed as follows:
\begin{equation}
C_\text{surf} (t) = \frac{1}{N_\text{tot}} \sum_{i = 0}^{N_\text{tot}}\dfrac{\left< I_{\text{X}, i} (t) I_{\text{X}, i} (0) \right>}{\left< I_{\text{X}, i} (t) \right>},
\label{eq:GI}
\end{equation}
where $C_\text{surf} (0) = 1$, and $C_\text{surf} (+ \infty) = N_\text{surf} / N_\text{tot}$. The surface spectral density $J_\text{surf} (\omega)$ is obtained from the Fourier transform of $C_\text{surf} (t)$. By analogy with Eq.\eqref{eq:T1RT_bulk}, the surface contribution to the relaxation rate, $R_{1, \text{surf}} (\omega_0)$, is calculated as:
\begin{equation}
R_{1, \text{surf}} (\omega_0) \propto J_\text{surf} (\omega_0) + 4 J_\text{surf}  (2 \omega_0).
\label{eq:RI2}
\end{equation}

\section{$J_\text{surf}$ from first passage time}

\noindent Molecules adsorbing at the top surface (T) are explicitly differentiated from those adsorbing at the bottom surface (B). At the bottom wall, the single-molecule correlation function can be expressed as follows:
\begin{equation}
C_\text{B} (t) = \left< I_\text{B} (0) I_\text{B} (t) \right> / \left< I_\text{B} (0) \right>,
\label{eq:CAA}
\end{equation}
where $I_\text{B} = 1$ when the molecule is adsorbed at the bottom wall and $I_\text{B} = 0$ otherwise. The expression $\left< I_\text{B} (0) \right> = \left< I_\text{B} (t) \right>$ represents the probability of the molecule being adsorbed at the bottom wall. The same equations can be written for the top wall (T).

The Laplace transform $\tilde{C}_\text{B} (s)$ of $C_\text{B} (t)$ is given by \cite{gravelle19a}
\begin{equation}
\tilde{C}_\text{B} (s) = \dfrac{\tilde{Q}_\text{B} (s) \tilde{J}_\text{B}^* (s)}
{1 - \tilde{\psi}_\text{B} (s) \tilde{J}_\text{B}^* (s) \tilde{J}_\text{B $\to$ T} (s) \tilde{J}_\text{T}^* (s) \tilde{\psi}_\text{T} (s) \tilde{J}_\text{T $\to$ B} (s)},
\label{eq:C_BB}
\end{equation}
where $\tilde{C}_\text{B} (s)$ explicitly accounts for all desorption and re-adsorption events that the molecule undergoes over time. $\tilde{\psi}_\text{B} (s)$ in Eq.\,\eqref{eq:C_BB} is the Laplace transform of the probability function $\psi_\text{B} (t)$ for a molecule adsorbed at the bottom wall to desorb at time $t$, assuming that it was adsorbed at time $t = 0$. Here, $\psi_\text{B} (t)$ follows an exponential distribution, and its Laplace transform reads 
\begin{equation}
\tilde{\psi}_\text{B} (s) = \dfrac{\lambda_\text{desorb}}{\lambda_\text{desorb} + s}.
\end{equation}
Assuming that molecules desorb from the top and bottom walls with the same rate $\lambda_\text{desorb}$, the same equation can be written for $\tilde{\psi}_\text{T} (s)$. $\tilde{Q}_\text{B} (s)$ in Eq.\,\eqref{eq:C_BB} is the Laplace transform of the survival probability $Q_\text{B} (t) = \int_t^\infty \psi_\text{B} (t') \mathrm{d}t'$, and reads
\begin{equation}
\tilde{Q}_\text{B} (s) = (\lambda_\text{desorb} + s )^{-1}.
\end{equation}

$\tilde{J}_\text{B}^* (s)$ in Eq.\,\eqref{eq:C_BB} is the renormalized first return probability that accounts only for re-adsorption events onto the bottom wall, i.e. molecules that do not adsorb onto the top wall after desorbing from the bottom wall. It is given by
\begin{equation}
\tilde{J}_\text{B}^* (s) = [1 - \tilde{\psi}_\text{B} (s) \tilde{J}_\text{B $\to$ B} (s)]^{-1}.
\end{equation}

Finally, $\tilde{J}_\text{B $\to$ B}$ is the first-return distribution for a molecule leaving the surface (B) before returning to the same surface, while $\tilde{J}_\text{B $\to$ T}$ the first-return distributions for a molecule leaving surface B before  adsorbing to the top surface (T). Both $\tilde{J}_\text{B $\to$ B}$ and $\tilde{J}_\text{B $\to$ T}$ are calculated by solving the 1D diffusion equation \cite{gravelle19a}
\begin{equation}
\partial_t {G} (z, t | z_0) = D \partial_z^2 {G} (z, t | z_0),
\label{eq:diffusion}
\end{equation}
where ${G}$ is the Green function. Equation~\eqref{eq:diffusion} can be solved using the Laplace transformed equation
\begin{equation}
(s - D \partial_z^2) {\tilde G} (z, s | z_0) = \delta (z - z_0),
\label{eq:diffusion-laplace}
\end{equation}
where the initial condition ${\tilde G} (z, t=0 | z_0) = \delta (z - z_0)$ is used,
and where ${\tilde G} (z, s | z_0)$ is the Laplace transform of ${G} (z, t | z_0)$.

The general solution of Eq.\,\eqref{eq:diffusion-laplace} is given by:
\begin{equation}
\tilde {G} (z, s | z_0) = \left\{
    \begin{array}{ll}
        a \mathrm{e}^{-z \sqrt{s / D}} + b \mathrm{e}^{z \sqrt{s / D}} ~ \text{for} ~ 0 < z < z_0 < H, \\
        c \mathrm{e}^{-z \sqrt{s / D}} + d \mathrm{e}^{z \sqrt{s / D}} ~ \text{for} ~ 0 < z_0 < z < H.
    \end{array}
\right.
\end{equation}

Here, for simplicity, the bottom wall is assumed to be located at $z=0$ and the top wall at $z=H$. The coefficients $a$, $b$, $c$, and $d$ can be determined using four boundary conditions: a continuity equation at $z=z_0$, an initial condition, and two surface reaction boundary conditions (one for each surface). The reactive boundary condition at the bottom wall (at $z=0$) is expressed as:
\begin{equation}
D \partial_z \tilde{G} (0, s | z_0) = k \tilde{G} (0, s | z_0),
\label{eq:reactive}
\end{equation}
where $k$ is a phenomenological rate constant (in m/s), defined as
\begin{equation}
k = \dfrac{H N_\text{surf}}{N_\text{bulk}} \lambda_\text{desorb}.
\label{eq:k}
\end{equation}
The first-return distribution for a molecule leaving the bottom surface (B) before returning to the same surface is given by \cite{gravelle19a}
\begin{equation}
\tilde{J}_\text{B $\to$ B} (s) = k^* \dfrac{  1 - k^* + (1+k^*) \mathrm{e}^{2 H^*}}
{ (1 + k^*)^2 \mathrm{e}^{2 H^*} - (1 - k^*)^2},
\end{equation}
using $k^* = k / \sqrt{s D}$ and $H^* = H \sqrt{s / D}$. Similarly, the first-return distribution for a molecule leaving the bottom surface (B) before adsorbing at the opposite top surface (T) is given by:
\begin{equation}
\tilde{J}_\text{B $\to$ T} (s) = \dfrac{ 2 k^* \mathrm{e}^{H^*}} {(1 + k^*)^2 \mathrm{e}^{2 H^*} - (1 - k^*)^2 }.
\label{eq:JAB}
\end{equation}

Finally, the surface spectrum is calculated as follows:
\begin{equation}
J_\text{surf} (\omega) = \tilde C_\text{B} (i \omega) + \tilde C_\text{B} (- i \omega),
\label{eq:Jsurf_Netz}
\end{equation}
where $k$ is determined using Eq.\,\eqref{eq:k} with $N_\text{surf}$, $N_\text{bulk}$, and $\lambda_\text{desorb}$ measured directly from MD simulations.

\section*{Choice of surface population definition}

\noindent By default, the position $z_\rho$ of the first density depth was utilized to differentiate the surface from the bulk populations. Other definitions have been used in the past, such as the position between the second and third density layers \cite{levitz13a}. This alternative choice could be judicious for the most hydrophilic surfaces considered here, where density oscillations extend over multiple molecule layers (Fig.\,\ref{fig:system}\,c of the main text). We tested the sensitivity of our result to the choice of surface population definition and found that the surface exchange statistics are qualitatively similar whether the first or the second-density depth is used (Fig.\,\ref{fig:S-sensitivity-1}-\ref{fig:S-sensitivity-2}). The main impacts of using two surface layers instead of one as a surface population are a decreased value for $\lambda_\text{desorb}$, which is expected due to the larger surface thickness $\delta$, as well as a lower overall sensitivity of surface exchange statistics in the properties of the interface (Fig.\,\ref{fig:S-sensitivity-3}). 

\subsection*{Use of ITIM for surface detection}

\noindent All results from the main text were obtained by separating the surface and the bulk populations using the position of the first density depth extracted from the density profile $\rho(z)$. Extracting $z_\rho$ may not always be straightforward, particularly for atomically rough surfaces or in the case where the structuring of the fluid is insufficient, as is the case for $\epsilon_\text{sl} = 0.1$\,kJ/mol. As an alternative, the ITIM algorithm of Pytim \cite{partay08c, sega18a} can be used to detect the so-called truly interfacial molecules, i.e. the molecules that are part of the first layer of fluid next to the interface. As a proof of concept, we detected the surface populations for both hydrophilic and hydrophobic surfaces using ITIM (Fig.\,\ref{fig:PITIM}). The results obtained with ITIM align well with the data extracted using the ensity depth $z_\rho$. We note however that, in cases where ITIM is applied, the probe sphere radius influences the detected surface population and must be chosen with care.

\end{document}